\documentclass[pra,reprint,superscriptaddress,amsmath,amssymb,aps,showpacs, floatfix]{revtex4-1}

\usepackage{graphicx}
\usepackage[utf8]{inputenc}
\usepackage{epstopdf}
\usepackage{booktabs}
\usepackage{color}
\usepackage{xcolor}
\usepackage[percent]{overpic}

\newcommand{\ba}{\begin{eqnarray}}
\newcommand{\ea}{\end{eqnarray}}
\newcommand{\be}{\begin{equation}}
\newcommand{\ee}{\end{equation}}
\newcommand{\bea}{\begin{eqnarray}}
\newcommand{\eea}{\end{eqnarray}}

\newcommand{\re}{\mathrm{Re}\,}
\newcommand{\im}{\mathrm{Im}\,}

\newcommand{\ket}[1]{\ensuremath{\vert #1 \rangle}}

\def\openone{\leavevmode\hbox{\small1\kern-3.3pt\normalsize1}}

\newcommand{\bracket}[2]{\ensuremath{\langle #1 \vert #2 \rangle}}

\usepackage{theorem}

\theorembodyfont{\rmfamily}

\def\openone{\leavevmode\hbox{\small1\kern-3.3pt\normalsize1}}

\begin{document}
\title{Dressing the chopped-random-basis optimization: a bandwidth-limited access to the trap-free landscape}
\author{N. Rach}
\affiliation{Institute for Complex Quantum Systems \& Center for Integrated Quantum Science and Technology, University of Ulm, Albert-Einstein-Allee 11, D-89069 Ulm, Germany}
\author{M. M. Müller}
\affiliation{Institute for Complex Quantum Systems \& Center for Integrated Quantum Science and Technology, University of Ulm, Albert-Einstein-Allee 11, D-89069 Ulm, Germany}
\affiliation{LENS \& Dipartimento di Fisica e Astronomia, Università di Firenze, via G. Sansone 1, I-50019 Sesto Fiorentino, Italy}
\author{T. Calarco}
\affiliation{Institute for Complex Quantum Systems \& Center for Integrated Quantum Science and Technology, University of Ulm, Albert-Einstein-Allee 11, D-89069 Ulm, Germany}
\author{S. Montangero}
\affiliation{Institute for Complex Quantum Systems \& Center for Integrated Quantum Science and Technology, University of Ulm, Albert-Einstein-Allee 11, D-89069 Ulm, Germany}
\date{\today}
%03.65.Aa 	Quantum systems with finite Hilbert space
%02.30.Yy 	Control theory
%02.60.Pn Numerical optimization
%03.67.-a 	Quantum information
%
%
%
%
%

\begin{abstract}
In quantum optimal control theory the success of an optimization algorithm is highly influenced by how the figure of merit to be optimized behaves as a function of the control field, i.e. by the control landscape. Constraints on the control field introduce local minima in the landscape --false traps-- which might prevent an efficient solution of the optimal control problem. Rabitz et al. [Science 303, 1998 (2004)] showed that local minima occur only rarely for unconstrained optimization. Here, we extend this result 
to the case of bandwidth-limited control pulses showing that in this case one can eliminate the false traps arising from the constraint. 
Based on this theoretical understanding, we modify the Chopped Random Basis (CRAB) optimal control algorithm and show that this development exploits the advantages of both (unconstrained) gradient algorithms and of truncated basis methods, allowing to always follow the gradient of the unconstrained landscape by bandwidth-limited control functions. We study the effects of additional constraints and show that for reasonable constraints the convergence properties are still maintained. Finally, we numerically show that this approach saturates the theoretical bound on the minimal bandwidth of the control needed to optimally drive the system.
\end{abstract}

\maketitle
The ability of achieving a desired transformation of a quantum system lies at the heart of the success of experiments in cold atoms~\cite{BEC,opticallattice}, quantum optics~\cite{light-harvesting, photon, rydberg}, condensed matter~\cite{circuitQED}, and in quantum technologies~\cite{quantumtechnology}. Together with the fast development of these fields and the increasing complexity of the experiments, 
to develop efficient protocols it is often necessary to automatize the optimization process if not the whole 
development of the experimental sequence~\cite{Blatt}.
One possible way to perform such an optimization is by means of quantum optimal control, an approach that has proven to be very successful in solving this class of problems~\cite{Brif}.

The success of gradient methods to find global solutions in quantum control problems~\cite{Grape,Krotov,dalessandro} is largely due to the fact that local minima are very rare in the control landscapes of a large class of systems~\cite{Rabitz,Ho2006,Fouquieres,Moore2010,Wu,PechenRabitzDiscussion,Pechen2012,Pechen2014,Riviello}. However, to face the new challenges posed by the recent advancements in quantum science, gradient methods might be not the best option since to numerically calculate the gradient of the control objective might be quite inefficient. Moreover in an increasing number of interesting applications, the control objective does not allow for an analytical calculation of the gradient~\cite{ESU,PerfEnt}. 
The CRAB optimal control algorithm operates with an expansion of the control field onto a truncated basis and a direct optimization of the coefficients of the expansion by means of gradient-free
 minimization~\cite{Doria,Caneva}. These characteristics allow for the solution of optimal control problems involving many-body quantum systems~\cite{ESU,Caneva2}, as well as in the presence of long-range interactions ~\cite{Mueller2013}, bandwidth-limited control according to experimental constraints~\cite{Scheuer,Rosi,vanFrank}, and highly nonlinear functionals where the gradient can not be calculated~\cite{ESU,PerfEnt}.
However, the CRAB optimization does not necessarily fulfill the condition under which local minima in the optimization occur only rarely as it is by construction bandwidth-limited~\cite{Rabitz,Ho2006,Fouquieres,Moore2010,Wu,PechenRabitzDiscussion,Riviello}. 

In this paper we extend the results presented by Rabitz and coworkers~\cite{Rabitz} and show that bandwidth-limited optimal control can be made virtually free of local minima as in the case of unconstrained control. Moreover, we numerically show that the global minima one always reaches correspond to an optimal solution if the bandwidth satisfy the theoretical bound given in Reference~\cite{LloydMontangero}. In particular, we present an extension of the CRAB algorithm -- the ``dressed CRAB'' (dCRAB) -- that keeps the benefits of the original algorithm and comes with the additional property of guaranteed convergence to the global optimum in the cases where this is guaranteed also for gradient methods. We test the two versions of the algorithm by optimizing a state transfer for different instances of a random spin Hamiltonian. Being $M$ the dimension of the Hilbert space the problem is defined on, for the standard CRAB algorithm false traps occur when the heuristic ``$2 M-1$-rule'' for the number of required control coefficients is violated~\cite{MooreRabitz}. On the contrary, using the dCRAB presented hereafter, all false traps are removed regardless of the number of control coefficients, resulting also in a faster convergence to the global optimum. We present a theoretical explanation supporting these findings and showing that one can construct a set of random basis functions which follow the instantaneous gradient of the control landscape. Finally, we examine the behavior of dCRAB in the presence of additional constraints on the bandwidth and amplitude of the pulse, and show that the algorithm is well-behaving as false traps appear only in the presence of strong constraints. In the following, for simplicity we will focus on state to state transfer of pure states, however the theoretical arguments are valid in the general scenario of unitary gate generation and mixed-states optimal control. 

The structure of this paper is as follows: we first review the standard CRAB algorithm and introduce its extension, dCRAB~\cite{dressedCRAB}. In section II we present the theoretical basis of dCRAB, while in
section III we show that dCRAB follows the instantaneous gradient of the landscape. Finally, in Section IV we apply both CRAB and dCRAB to control problems with random spin Hamiltonians and evaluate their performance with respect to success probability, computational effort and behavior under constraints.

\section{From CRAB to dressed CRAB}\label{sec:CRAB}
In the following, we consider the optimization of a state-to-state transfer problem where the figure of merit is the overlap of the final state after the evolution with the target state
\begin{eqnarray}\label{eq:fidelity}
F(|\psi(T)\rangle)=|\langle\zeta|\psi(T)\rangle|^2\,.
\end{eqnarray}
Here $|\zeta\rangle$ is the target state and the time evolution is given by the Schr\"odinger equation
\begin{align}
i\frac{\partial}{\partial t}|\psi(t)\rangle=\left(H_0+f(t)H_1\right)|\psi(t)\rangle\nonumber\\
|\psi(0)\rangle=|\xi\rangle
\end{align}
with initial state $|\xi\rangle$ and control function $f(t)$ is determined by the optimization algorithm in order to maximize the figure of merit.

The CRAB algorithm~\cite{Doria,Caneva} builds on the fact that in most scenarios the resources available to solve an 
optimal control problem -- such as time, energy and bandwidth -- are limited: in particular, as the set of practically accessible
wave functions is usually limited, one can show that also the control bandwidth of the control field can be upper bounded~\cite{LloydMontangero}.    
This bound has very important consequences as optimal control problems can be practically solved by exploring a small subset of the 
a priori infinite dimensional search space of functions. That is, one can expand the control field in a truncated basis
\begin{align}\label{eq:CRAB1}
f(t)=\sum_{i=1}^{N_C}c_{i}f_{i}(t)
\end{align}
and the optimization can then be performed on this subspace of {\it small} dimension, resulting in an optimal set of coefficients 
$c_{i}, i=1, \cdots, N_C$. 
The optimization can be performed by standard tools, e.g. by the Nelder-Mead simplex algorithm that does not rely on gradients~\cite{NelderMead}.
A standard choice for the basis functions $f_i(t)$ are trigonometric functions, often multiplied by a shape function $1/\gamma(t)$ that fixes the pulse boundary conditions and a guess pulse~\cite{Doria,Caneva}. 

Due to the restriction of the search basis to $N_C$ dimensions given by the CRAB expansion in Eq.~\eqref{eq:CRAB1}, the algorithm might converge to a non-optimal fixed point, i.e. the algorithm is trapped in a local minimum arising due to the constraint - a so-called false trap. To overcome this problem, and escape from these false traps, we show in the next section that one can start from the non-optimal fixed point a new CRAB optimization with a new random basis and new coefficients. 
This is done in an iterative way so that in the $j$-th super-iteration 
one optimizes the coefficients $c_i^j$ of
\begin{eqnarray}\label{eq:dCRAB}
 f^j(t)=f^{j-1}(t)+\sum_{i=1}^{N_C}c_i^j f_i^j(t)\,,
\end{eqnarray}
where $f_i^j(t)$ are new randomly chosen basis functions, for example sine or cosine functions with random frequencies within some interval $[0,\omega_{\mathrm{max}}]$. 
As a consequence, in each super-iteration the old pulse is dressed with new search directions and we call this procedure dressed Chopped Random Basis (dCRAB) algorithm~\cite{dressedCRAB}.
This updating of the search directions can also be understood as an extension of Powell's method \cite{Powell} to an infinite dimensional search space.
In the following sections we give a theoretical explanation why this is a substantial improvement of the algorithm by analyzing how it influences the constrained control landscape and we demonstrate it numerically by applying it to a model with typical properties.

\begin{figure}
\includegraphics[width=0.5\textwidth]{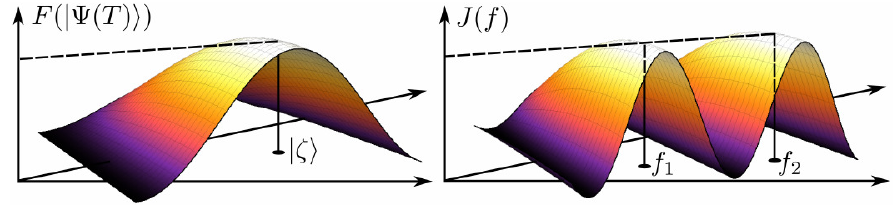}
\caption{(Color online) Schematic view on the control landscape $J(f)=F(|\psi(T)\rangle)$. While $F$ has a strict global maximum point at $|\zeta\rangle$ (left panel), this state can be reached by different control functions $f_1$ and $f_2$ corresponding to multiple global maximum points in the landscape $J$ (right panel).}\label{fig:landscape}
\end{figure}

\section{Control Landscapes and CRAB}\label{sec:landscapes}
In this and in the following section, we review the theory of control landscapes and how it can enlighten the reasons why optimal control algorithms converge to the optimal solution or become trapped~\cite{Brif,Rabitz}. We first review a general perturbative analyses of control landscapes presented in \cite{Wu}, specifically its gradients and critical points, valid independently from the particular optimization algorithm employed to find the optimal driving field; and then we specify it to the CRAB approach we are focusing here. In particular, we focus on why CRAB can be trapped whereas dCRAB – like gradient methods – cannot be trapped, as we will show also in the next section by means of some numerical examples.

Given a control problem with control $f(t)$, the control landscape is the functional $J(f)$ with
\begin{align}\label{eq:controllandscape}
 J(f)=F(|\psi(T)\rangle)\,
\end{align}
where \ket{\psi(T)} is the final state resulting from time evolution of the initial state \ket{\xi} with the given control $f(t)$ and $F$ is a figure of merit quantifying the quality of the process~\cite{Brif,Rabitz}.
In other words, the control landscape is the functional expressing the figure of merit of the process as a function of the control field. In the following we use the state fidelity given by Eq.~\eqref{eq:fidelity} as a figure of merit. Notice that the control landscape can be seen either as a function of the final state $| \psi(T)\rangle$ or as a functional of the control field $f(t)$: in the first case the function has by definition a single optimal point ($|\psi(T)\rangle=|\zeta\rangle$, up to a global phase), while in the latter there might exist different control pulses leading to the same maximum value, as  
sketched in Fig.~\ref{fig:landscape}.

Optimization usually leads to so-called critical points of the landscape, that is to pulse shapes fulfilling the condition
\begin{align}\label{eq:criticalpoint-chainrule}
 \delta J=\langle \nabla F(\psi(T))|\delta \psi(T)\rangle=0\quad\forall\;\;\delta f\,,
\end{align}
i.e. a vanishing variation of the functional $J$ for a variation of the control $f$. Using the chain rule one can show that this variation consists of two parts: the gradient of the fidelity as a function of the final state
$\nabla F/ \delta \psi(T)$, and the variation of the final state as a result of the variation of the control $\delta \psi(T)/\delta f$. 
For all common choices of the figure of merit a vanishing gradient $\nabla F(\psi(T))=0$ corresponds to the global maximum, global minimum or a saddle point~\cite{Brif,Rabitz,Wu}; 
more specifically if $F(|\psi(T)\rangle)$ is the state overlap (eq. \eqref{eq:fidelity}) there are no saddle points since $\nabla F=0$ corresponds to $F=0$ (the global minimum) or $F=1$ (the global maximum). Less intuitive is the role played by the second term, 
which we analyze in the following, to understand under which conditions $\delta J=0$ implies  $\nabla F(\psi(T))=0$.

Critical points can be classified as singular or regular critical points, where regular
means that for every $|\delta\psi(T)\rangle$ in the Hilbert space of the problem there is a
change in the control $\delta f$ that generates it, while on the contrary for singular points this is not the case~\cite{Brif,Rabitz,Wu}.
As a direct consequence, for regular critical points we have
\begin{align}
 \delta J=\langle \nabla F(\psi(T))|\delta \psi(T)\rangle&=0\quad\forall\;\;\delta f\nonumber\\
 \Rightarrow\delta J=\langle \nabla F(\psi(T))|\delta \psi(T)\rangle&=0\quad\forall\;\;\delta\psi(T)\nonumber\\
 \Rightarrow\nabla F(\psi(T))&=0
\end{align}
as in this case the gradient of the fidelity is orthogonal to the whole Hilbert space. This means that a regular point is not a trap. That is, all traps have to be singular critical points (although the reverse is not necessarily true). 
However, at least for controllable systems, all known traps occur at constant control fields~\cite{PechenRabitzDiscussion,Fouquieres}, often at $f=0$, and numerical evidence hints to the fact that 
the singular points at non-constant control seem to be no traps~\cite{Moore2010,Wu,Riviello}. That is, it is commonly accepted that unconstrained control landscapes for all practical purposes have no traps, although a rigorous proof has been given only for two specific situations, the Landau-Zener system \cite{Pechen2012} and the transmission of a wave package over a potential barrier \cite{Pechen2014}.

Now we discuss the implications of these results on the CRAB and dCRAB optimal control success rate. As a starting point, we recall that the previous statements assume that the control $f$ is a general 
function in $L_2$ whereas for CRAB we are much more restricted by the expansion into the truncated basis. This means that one might find false traps in the CRAB control landscape, i.e.
encounter points where $\delta J$ vanishes for all variations $\delta f$ allowed by the truncated basis expansion of Eq.~\eqref{eq:CRAB1}, but not for all variations of the unconstrained control space:
\begin{align}\label{eq:pseudocriticalpoint}
 \delta J=0\;\forall\;\delta f=\sum_{i=1}^{N_c}f_i(t)\delta c_i\quad\text{but}\nonumber\\
 \exists\;\delta f\;:\;f+\delta f\in\;L_2\;,\;\delta J\neq 0\,.
\end{align}
These points are called false traps~\cite{MooreRabitz} as they arise only artificially from the choice of the basis and their influence can hinder convergence of the algorithm~\cite{MooreRabitz,Riviello2}. 

Here, we show how these false traps for CRAB are removed by the super-iterations of dCRAB:
From equations (\ref{eq:criticalpoint-chainrule}) and (\ref{eq:pseudocriticalpoint}) together with
\begin{align}\label{eq:gradF}
 \langle\nabla F(|\psi(T)\rangle)|\,\cdot\,\rangle = 2\re(\bracket{\psi(T)}{\zeta}\bracket{\zeta}{\,\cdot\,})
\end{align}
it follows that, for a false trap the gradient has to be $\bracket{\nabla F(\psi(T))}{\,\cdot\,}=\re\bracket{\phi_T}{\,\cdot\,}$ for some non-zero vector $\ket{\phi_T}$.
A perturbative treatment yields~\cite{Wu}:
\begin{align}
 |\delta\psi(T)\rangle=-i U(T)\int_0^T U^\dagger (t)H_1U(t)|\xi\rangle\delta f(t)\mathrm{d}t\,.
\end{align}
This results in the expression for the overlap of the state update and the gradient
\begin{align}
\delta J= \re\langle\phi_T|\delta\psi(T) \rangle
 =\int_0^T k(t)\delta f(t)\mathrm{d}t\\
 k(t)=-\im\langle \phi_T|U(T) U^\dagger (t)H_1U(t)|\xi\rangle
\end{align}
where $k$ is a continuous function and $k\neq0$ since $k=0$ would violate Eq.~(\ref{eq:pseudocriticalpoint}).
If we now choose $\delta f(t)=\sin(\omega_r t)\delta c$ with a new random frequency $\omega_r$  we get
\begin{align}\label{eq:almostsureoverlap}
\int_0^T k(t)\delta f(t)\mathrm{d}t\neq 0\,
\end{align}
almost surely (i.e. the integral vanishes only on a null set of the probability measure).
If we now perturb our control field $f$ by this new frequency contribution (this is the new super-iteration of dCRAB) we find
\begin{align}\label{eq:outofartificaltrap}
 \delta J =\langle \nabla F(\psi(T))|\delta \psi(T)\rangle\neq 0
\end{align}
and thus we have removed the false trap. In conclusion, the recipe to escape from a false trap is simply to add a new random frequency term to the CRAB expansion once we are in the false trap. This can be done also without increasing the total number of coefficients since when one is at the bottom of a false trap for a given set of basis functions and coefficients there is no more use in varying the old coefficients, and they can be kept at their value. We thus use the pulse $f$ leading to the false trap as a guess pulse for the next super-iteration of dCRAB (as explained in the previous section, see  Eq.~(\ref{eq:dCRAB})) that opens the false trap.

\section{Instantaneous basis functions for pulse update}\label{sec:basisfunctions}
In this section we show that by choosing enough new frequencies in a single super-iteration of dCRAB we can follow approximately the instantaneous gradient of the control landscape with a bandwidth-limited pulse update.
In the previous section we chose a new basis function so that its scalar product with
\begin{align}
 k(t)= -\im\langle \phi_T|U(T) U^\dagger (t)H_1U(t)|\xi\rangle
\end{align}
would be finite.
It is worth noting that this function $k(t)$ is exactly the update direction used in gradient algorithms with the time-evolved initial state
\begin{eqnarray}
 \ket{\psi(t)}=U(t)\ket{\xi}
\end{eqnarray}
and the adjoint state
\begin{eqnarray}
 \ket{\chi(t)}=U(t)U^{\dagger}(T)\ket{\nabla F(\psi(T))}\,.
\end{eqnarray}
Together with $H_1=\frac{\delta(H_0+fH_1)}{\delta f}=\frac{\delta H}{\delta f}$ we get the well-known
\begin{eqnarray}
 k(t)= -\im\bigg\langle \chi(t)\bigg\vert\frac{\delta H}{\delta f}(t)\bigg\vert\psi(t)\bigg\rangle\,.
\end{eqnarray}

To follow the gradient of the control landscape only a limited freedom for the pulse update is needed. Hsieh et al.~\cite{Hsieh} derived for a state to state transfer in an $M$-level system a set of $2M-2$ functions (for each iteration step) that span the pulse update $k(t)$. This simply reflects the fact that the state update generated by the pulse update $\delta f$, i.e. $|\delta\psi(T)\rangle$, is a $M$ dimensional complex vector ($2M$ real coefficients) with norm $1$ and irrelevant global phase (thus $2M-2$ real coefficients).

We can generate something very similar by choosing $2M-1$ random frequencies (we do not fix the global phase here) and corresponding pulse updates $\delta f_n(t)=\sin(\omega_n t)\delta c_n$ ($n=1,\dots 2M-1$) with real coefficients $c_n$. Let us here assume that $L_2$ pulses can generate $|\delta\psi(T)\rangle$ out of the whole tangential space of the unity sphere of $\mathbb{C}^M$ (i.e. regularity of the control landscape at the point of consideration which is the general case for trap free landscapes). The state update generated by each pulse update in a perturbative regime reads~\cite{Wu}:
\begin{align}
 \ket{\delta \psi_n(T)}=-i U(T)\int_0^T U^\dagger(t) H_1 U(t)|\xi\rangle\delta f_n (t) dt\,.
\end{align}

We now show that for every vector $|\delta\phi\rangle$ out of this unit sphere there is a set of real coefficients $\alpha_n$ so that $|\delta \phi\rangle=\sum_{n=1}^{2M-1}\alpha_n |\delta\psi_n(T)\rangle$.
In order to do so we identify $|\delta \phi\rangle$ and $|\delta\psi_n(T)\rangle$ as elements of a $2M-1$ dimensional real vector space with scalar product $\re\bracket{v}{w}$ and show that the $|\delta\psi_n(T)\rangle$ are a basis of this space. Let us thus consider $P_n=\mathrm{span}\{|\delta \psi_1(T)\rangle, \dots , |\delta\psi_n(T)\rangle\}$. We will show by induction that $P_n$ has dimension $n$. For $n=1$ this is trivial. If now $P_n$ has dimension $n$ we have to show that $|\delta\psi_{n+1}(T)\rangle$ is not orthogonal to $P_n^\bot$. Let thus be $v\in P_n^\bot$. The overlap of $|\delta\psi_{n+1}(T)\rangle$ and $|v\rangle$ is
\begin{align}
 \re\langle v|\delta\psi_{n+1}(T) \rangle=\int_0^T l(t)\delta f_{n+1}(t)\mathrm{d}t\\
 l(t)=-\im \langle v|U(T) U^\dagger (t)H_1U(t)|\xi\rangle\nonumber
\end{align}
The kernel $l(t)$ is continuous and if $l\neq 0$ the overlap is nonzero almost surely.
However, $l=0$ $\Leftrightarrow$ $|v\rangle=\alpha |\psi(T)\rangle$ ($\alpha \in \mathbb{R}$) \cite{footnote:l0} and thus $|v\rangle$ is not in the tangential space.
This proves that the $|\delta\psi_n(T)\rangle$ as span the whole $2M-1$ dimensional real vector space, or in other words the $|\delta\psi_n(T)\rangle$ with real coefficients span the whole tangential space of the unit sphere of $\mathbb{C}^M$ (almost surely).

We can also orthogonalize the state updates in a Gram Schmidt way to obtain the orthogonal basis $|\delta\tilde\psi_n(T)\rangle$. This translates into a change of the pulse updates via
\begin{align}
 \delta \tilde{f}_n(t)=\delta f_n(t) -\sum_{k=1}^{n-1} \langle\delta \tilde{\psi_k}(T)|\delta\psi_n(T)\rangle\delta\tilde{f}^k(t)\,,
\end{align}
with $|\delta \tilde{\psi_n}(T)\rangle$ generated by $\delta \tilde{f}_n(t)$. Each direction $\ket{\delta \phi}$ is then a linear superposition of the basis states
\begin{eqnarray}
 \ket{\delta\phi}=\sum_{n=1}^{2M-1}\alpha_n\ket{\delta\tilde{\psi}_n(T)}
\end{eqnarray}
and can thus be generated by a superposition of the corresponding pulses
\begin{eqnarray}
 \ket{\delta\phi}=-i\int_0^T U(T)U^{\dagger}(t)H_1 U(t)\ket{\xi}\sum_{n=1}^{2M-1}\alpha_n \delta\tilde{f}_n(t) dt\,.\nonumber\\
\end{eqnarray}
In particular this proves that with $2M-1$ random instantaneous basis functions ($\delta f_n(t)$ oder $\delta \tilde{f}_n(t)$ for the ``orthogonal'' basis) we can follow the instantaneous gradient of the control landscape by suitable coefficients. 
We stress that, unlike in the case of a standard gradient method, this is done by bandwidth-limited pulses which can be comfortably 
adapted to typical experimental constraints such as bandwidth-limited control electronics.

This result demonstrates that after a change of the truncated basis in the function space the search in the state space is quasi-local. We can use this in the numerical part by initializing a small simplex after the basis change to ensure local search. The orthogonalization procedure could also help to find a better basis for the CRAB (re)start at the cost of $M(M-1)/2$ additional function evaluations.

\section{Numerical experiment}\label{sec:systemsandresults}
In this section we show that the dCRAB algorithm is indeed capable of escaping from false traps in random optimal control problems of increasing complexity. 
We consider a spin Hamiltonian
\begin{equation}\label{eq:hamiltonian}
 H=\sum_{i=1}^N\alpha_i\sigma_i^x+\beta_i\sigma_i^z + f(t)\sum_{i=1}^{N-1}\sigma_i^z\sigma_{i+1}^z
\end{equation}
with random coefficients $\alpha_i$, $\beta_i\in\,[0,1]$ and a control field $f(t)$ tuning the interaction. The random coefficients lift the symmetry of the system and make it controllable \cite{Altafini}.

We investigate the control resources needed to drive the system from a random initial state $|\xi\rangle$ to a random target state $|\zeta\rangle$ within a fixed time interval $[0,T]$. We measure the fidelity of this state-to-state transfer by Eq.~\eqref{eq:fidelity}. We investigate different system sizes and different operation time intervals, analyze the occurrence of false traps in the landscape by several trials of optimization of the fidelity $F$ for different random instances of $\ket{\xi}$ and $\ket{\zeta}$ (uniformly distributed over the unit sphere), and measure the success of an optimization by a certain threshold $\eta=10^{-3}$ for the residual error $\varepsilon=1-F$. We count an optimization trial as a success if $\varepsilon<\eta$ after convergence of the optimization or after a certain maximum number of function evaluations. The success rate of the optimization depends on the presence or absence of false traps in the landscape. To detect them we optimize ten different sets of random instances of the Hamiltionian and of the initial and goal state;  
for each instance we use ten different random starting points and random sets of frequencies to perform the CRAB optimization.
The frequencies of the basis functions (both for CRAB and dCRAB) are chosen randomly out of an interval $[0,\omega_{max}]$: 
 we choose $T \omega_{\mathrm{max}}/ (2 \pi) = 8, 20, 40$ for $N=2,3,4$ respectively.
 \begin{figure}[t]
\includegraphics[width=0.47\textwidth]{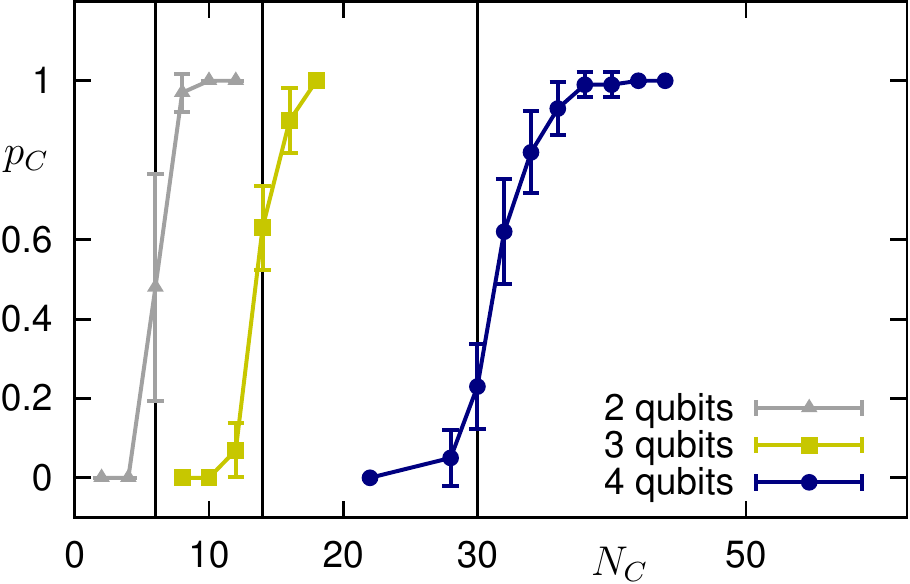}
\caption{(Color online) Success probability $p_C$ as a function of the number $N_C$ of basis functions in the CRAB expansion. As we increase the size of the function space false traps are removed. The symbols are the mean values over 10 different random pairs of final and initial states $|\xi\rangle$ and $|\zeta\rangle$ with 10 different starting points and frequencies each. The error bars show the standard deviation over the different final and initial states. The total time was set to $T=6\pi,10\pi,16\pi$ and the allowed bandwidth to $T \omega_{\mathrm{max}}/ (2 \pi) = 8, 20, 40$ for $N=2,3,4$ qubits respectively. An optimization was counted as success when the residual error $\varepsilon$ was smaller than $\eta=10^{-3}$. The black lines indicate the empirical ``$2\cdot 2^N-2$''~rule for the required number of coefficients $N_C$.
}\label{fig:nt-vs-freq}
\end{figure}

\begin{figure}[t]
\includegraphics[width=0.47\textwidth]{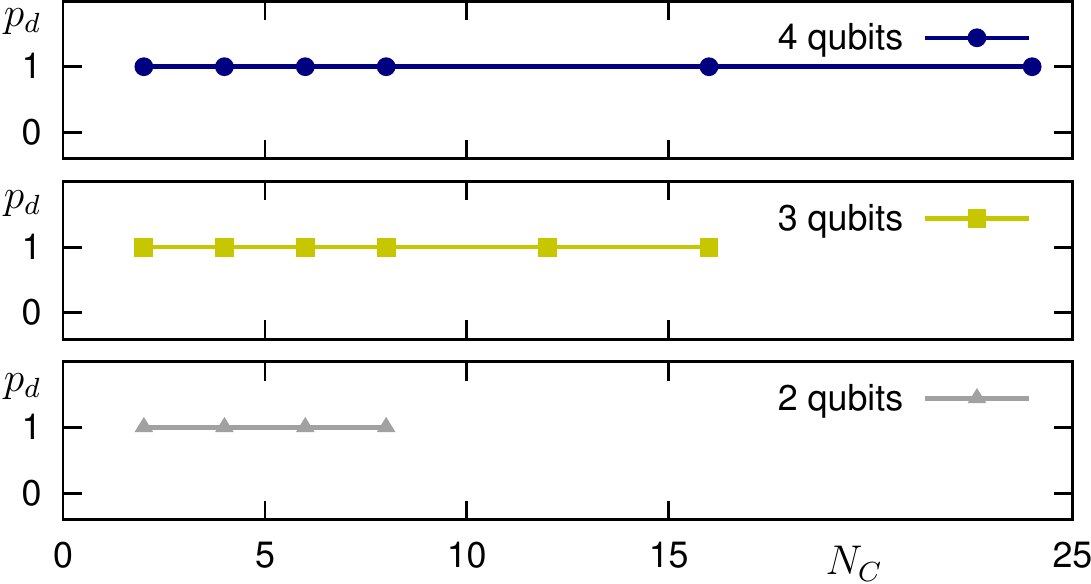}
\caption{(Color online) Success probability $p_d$ as a function of the number $N_C$ of basis functions in a single call of CRAB within the dCRAB super-iterations. Independently of $N_C$ no false traps are encountered and $p_d=1$. The total time, allowed bandwidth and error threshold are as in Fig.~\ref{fig:nt-vs-freq}.
}\label{fig:traps-removed}
\end{figure}

Fig.~\ref{fig:nt-vs-freq} reports the success probability $p_C$ for the standard CRAB optimization as a function of the number 
$N_C$ of coefficients of the truncated basis expansion in Eq.~\eqref{eq:CRAB1}: as can be clearly seen, for $N_C$ large enough no false traps are present, resulting in a success probability of one.  Here,  
by large enough we mean the empirical ``$N_C = 2\cdot 2^N -2$''~rule for unconstrained optimization, 
where the number of real coefficients equals the number of independent real entries in the state vector~\cite{MooreRabitz}.
The correspondent analysis for the dCRAB approach is reported in Fig.~\ref{fig:traps-removed}: the success probability $p_d$ is always one regardless of $N_C$, false traps are 
avoided and  the fidelity always exceeds the threshold.

\begin{figure}[h]
\includegraphics[width=0.47\textwidth]{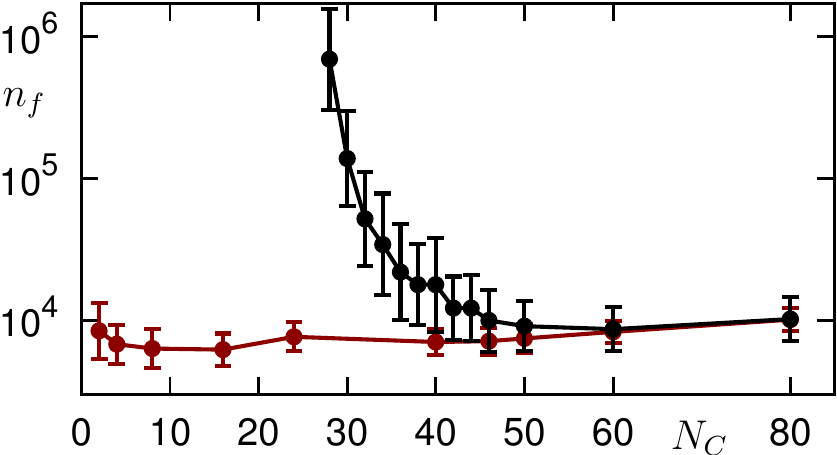}
\caption{(Color online) Number of function evaluations $n_{f}$ for dCRAB (red) and original CRAB (black) as a function of the number $N_C$ of basis functions involved in a single call of the respective algorithm.
The error bars show the logarithmic standard deviation. Optimization was stopped when the error $\varepsilon$ crossed the threshold $\eta=10^{-3}$. The total time was $T=16\pi$, the allowed bandwidth $T\omega_{max}/(2\pi)=40$ and the system size $N=4$ qubits. Similar results are obtained for other choices of the parameters. 
%--$T_2$
}\label{fig:aCRAB-4qbt}
\end{figure}

Despite the striking difference in terms of success probability, an important benchmark for any optimization method is given by the computational effort required to arrive at the optimal solution. Here we focus on the number of function evaluation needed to achieve the global optimum as this is practically the only difference between the two methods: Fig.~\ref{fig:aCRAB-4qbt} shows the number of function evaluations $n_{f}$ required by the two methods to exceed the threshold $F=1-\eta$ as a function of the coefficients $N_C$ in the CRAB expansion (in the case of dCRAB the coefficients of a single super-iteration).
All points consist of the average number of function evaluations of the successful runs divided by the respective success probability, that is the number of function evaluations that on average one has to do to solve the optimal control problem using one of the two methods. Note that the minimal effort does not follow the ``$2\cdot 2^N-2$''~rule of Fig.~\ref{fig:nt-vs-freq} for guaranteed convergence.
For CRAB the computational effort heavily depends on $N_C$, which can be problematic as the best choice of $N_C$ is not known in advance. For dCRAB instead there is only a minor dependance on $N_C$ in the order of magnitude of the error bars. Furthermore, Fig.~\ref{fig:aCRAB-4qbt} shows that even with the best choice of $N_C$ CRAB can not beat the performance of dCRAB.

\subsection{Constrained optimization}
Finally, we study how the convergence properties change in the presence of additional constraints -- like limited fluence and bandwidth typically present in experimental setups -- that violate the hypothesis of the analysis of control landscapes presented up to now. Indeed, in this scenario, false traps might be present which might change performance and convergence speed of the algorithms.
We consider separately two different kinds of constraints: bandwidth-limited control and limited pulse height. 

In the first case, we observe that Eq.~\eqref{eq:almostsureoverlap} still holds even if the new random frequency $\omega_r$ is chosen only within the limited bandwidth interval $[0,\omega_{\mathrm{max}}]$. We can then study the performance of the optimization as a function of $\omega_{\mathrm{max}}$, as done in the unconstrained case and compare the CRAB and dCRAB approaches.
\begin{figure}[t]
\includegraphics[width=0.47\textwidth]{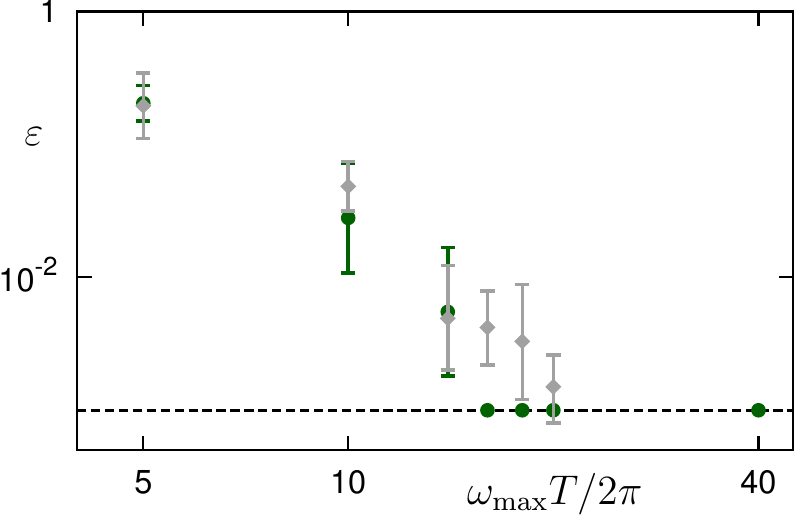}
\caption{(Color online) Infidelity $\varepsilon=1-F$ as a function of the limited bandwidth $\omega_{max}$ for $\ket{\xi} = \ket{0000}$ and $\ket{\zeta} = \ket{1111}$ in the system of $N=4$ qubits and total time $T=16\pi$ with fixed random coefficients $\alpha_i$, $\beta_i$ in the Hamiltonian (Eq. \eqref{eq:hamiltonian}). The grey diamonds show the infidelity $\varepsilon$ for CRAB ($N_C=40$), while the green circles show it for dCRAB ($N_C=\max \{2\omega_{\mathrm{max}}T/2\pi,40\}$). The black dashed line indicates the error threshold of $\eta=10^{-3}$.
The error bars report the logarithmic standard deviation.}\label{fig:bandwidth}
\end{figure}
The results are reported in Fig.~\ref{fig:bandwidth} where we show the optimal  infidelity $1-F$ reached from ten independent runs as a function of $\omega_{max}$. 
The optimal control problem is to perform the state transfer from $\ket{\xi} = \ket{0000}$ to $\ket{\zeta} = \ket{1111}$ given the Hamiltonian of Eq.~\eqref{eq:hamiltonian} for CRAB (grey diamonds) and dCRAB (green dots).  
One can see that for $\omega_{max} \geq 16 \cdot 2 \pi/T$ dCRAB succeeds with probability one, as 
all instances reached the optimal fidelity. Notice that this is less than half the bandwidth of the previous results reported in Figs.~\ref{fig:traps-removed} and~\ref{fig:aCRAB-4qbt}.
In addition, in a small intermediate regime around $\omega_{max} = 14 \cdot 2 \pi/T$, some optimizations succeed while others fail indicating the presence of false traps (note that the graph shows just the mean value of the infidelity $\varepsilon$ and the standard deviation); while 
for $\omega_{max} \leq 10 \cdot 2 \pi/T$ the final state cannot be reached anymore, indicating that the bandwidth is too small to achieve the desired result. 
In the case of CRAB, the three regimes are shifted toward larger frequencies. 
The lower bound observed in both cases is in agreement with an information theoretical argument given in~\cite{LloydMontangero}: to achieve full control over the system the inequality $\omega_{max}T\geq D$ has to be fulfilled, where $D=32$ is the dimension of the state space. This inequality basically says that the control has to contain enough information to distinguish the target state from the other reachable states and it yields 
$\omega_{max}\gtrsim 5 \cdot 2\pi/T$, the value of the bandwidth where the infidelity in Fig.~\ref{fig:bandwidth} starts to drop indicating that control over the system starts to be effective. 

\begin{figure}[t]
\includegraphics[width=0.47\textwidth]{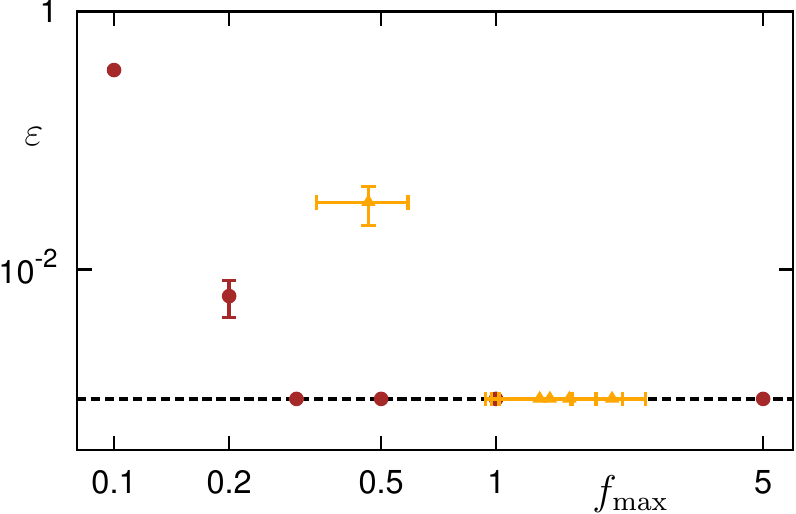}
\caption{(Color online) Infidelity $\varepsilon=1-F$ as a function of the limited pulse height $f_{max}$ for the transfer from $\ket{\xi} = \ket{00}$ to $\ket{\zeta} = \ket{11}$ in the 2-qubit system with constrained dCRAB. The black line indicates the error threshold of $\eta=10^{-3}$. The orange triangles are obtained with Lagrange multipliers (see Eq.~\eqref{eq:pulseheight-weight}), while the brown circles are obtained with a cut-off at $f_{max}$ (see Eq.~\eqref{eq:pulseheight-wall}). The error bars indicate the standard deviation over 10 different starting points of the optimization. The orange triangles have also errorbars in $f_{\mathrm{max}}$ since the Lagrange multiplier is not a hard wall and instead we plot the maximum absolute value the optimal pulse takes in each realization.
}\label{fig:pulseheight}
\end{figure}

We then perform a similar analysis for pulse height limited control with dCRAB, where we study two scenarios to include such a constraint. We first introduce a smooth constraint as usually done, that is a penalty on the pulse height so that the control objective becomes
\begin{eqnarray}\label{eq:pulseheight-weight}
J=F-\lambda \max_{t}|f(t)|\,.
\end{eqnarray}
As a second alternative, we limit the pulse height by a hard wall constraint, by using the update formula
\begin{eqnarray}\label{eq:pulseheight-wall}
\tilde{f}^j(t)=f^{j-1}(t)+\sum_{i=1}^{N_C}c_i^j f_i^j(t)\,,\\
 f^j(t)=\begin{cases}
         \tilde{f}^{j}(t)\qquad\text{if}\,\,|\tilde{f}^{j}(t)|<f_{\mathrm{max}}\\
         \mathrm{sgn}(f^j(t)) f_{\mathrm{max}}\qquad\text{otherwise}.
        \end{cases}
\end{eqnarray}
The results of these two procedures are reported in Fig.~\ref{fig:pulseheight} where we plot the infidelity $1-F$ as a function of the maximal pulse height $f_{max}$:
Clearly the optimization works better with the hard boundaries than with the Lagrange multiplier, as the fidelity threshold $1-\eta$ can be exceeded for about three times weaker pulses. This difference can be understood by the fact that the hard wall introduces pulses of higher bandwidth.
Compared to the unconstrained system we can decrease the maximal value of the pulse by a factor of fifteen, while keeping 100\,\% success probability. 
For even smaller cut-off $f_{\mathrm{max}}$ the small error bars indicate that optimization failure is most probably more due to a loss of controlability than due to false traps.

\section{Conclusions}
We have generalized the results presented in~\cite{Rabitz} to the case of bandwidth-limited control pulses, under the condition that the bandwidth satisfy the theoretical bound introduced in~\cite{LloydMontangero}.  

Thanks to this theoretical result, we have modified the CRAB optimization algorithm to efficiently combine the advantages of gradient methods with those of truncated basis methods. We showed that it is possible to exploit both the guaranteed convergence to the global optimum that gradient algorithms exhibit in the frequent case in which the kinematic ($F(|\psi(T)\rangle)$) and the dynamical ($J(f)$) landscapes are equivalent and the numerical gradient-free truncated basis approach. Moreover, we showed for two typical constraints, namely the limited fluence and the limited bandwidth constraints below the theoretical bound, that some of the convergence properties survive, if the constraints are carefully implemented in the optimization procedure. 

We expect that the presented results will allow to tackle in the near future both theoretically and experimentally even more complex many-body problems as done so far, as well as a broader variety of control objectives and constraints.

\begin{acknowledgments}
The authors acknowledge support from SFB/TRR 21, Q.com, the EU project SIQS, RYSQ and DIADEMS, and we thank the bwUniCluster~\cite{Cluster} for the computational resources.
\end{acknowledgments}


\begin{thebibliography}{10}
\bibitem{dressedCRAB}This is inspired but not to be confused with ``Dressed Crab'', i.e. a dish where the crab meat is presented in the crab's cleaned shell to enjoy the meat without any effort.
\bibitem{footnote:l0}To see this assume $|v\rangle\neq \alpha |\psi(T)\rangle$. Then by the regularity of the point there is a pulse update $\delta f$ so that the fidelity $\tilde{F}=\re\langle v|\psi(T)\rangle$ is improved by $\delta f$ as $\delta J=-\im \int_0^T\langle v|U(T) U^\dagger (t)H_1U(t)|\xi\rangle\delta f_{n+1}(t)\mathrm{d}t\neq 0$, i.e. $0\neq \int_0^T l(t)\delta f_{n+1}(t)\mathrm{d}t$ and thus $l\neq 0$. If instead $|v\rangle=\alpha |\psi(T)\rangle$ we have $l(t)=-\im \langle v|U(T) U^\dagger (t)H_1U(t)|\xi\rangle$, i.e. $l(t)=-\im \langle \xi| U^\dagger (t)H_1U(t)|\xi\rangle=0$.
\bibitem{Cluster}This work was performed on the computational resource bwUniCluster funded by the Ministry of Science, Research and Arts and the Universities of the State of Baden-Württemberg, Germany, within the framework program bwHPC.
\end{thebibliography}
\end{document}